# LMEEC: Layered Multi-Hop Energy Efficient Cluster-based Routing Protocol for Wireless Sensor Networks

Manel Khelifi, Assia Djabelkhir
ReSyD, Doctoral School in Computer Science
UAMB, Bejaia university, Algeria
manel.khelifi@gmail.com, assia.djabelkhir@gmail.com

*Abstract—* In this paper, we propose LMEEC, a cluster-based routing protocol with low energy consumption for wireless sensor networks. Our protocol is based on a strategy which aims to provide a more reasonable exploitation of the selected nodes (cluster-heads) energy. Simulation results show the effectiveness of LMEEC in decreasing the energy consumption, and in prolonging the network lifetime, compared to LEACH.

## I. INTRODUCTION

Recent advances in electronics and wireless communication technologies have enabled the development of miniature sensors at low cost. The small size of the sensors confines the embedded energy because they are alimented by non-rechargeable batteries, which are even not easily replaceable. Due to this constraint, a lot of work has been conducted to manage the sensors' energy consumption in order to extend their lifespan and thus, prolong the whole network lifetime. Moreover, in Wireless Sensor Network (WSN), the sensed data is communicated through the wireless medium to the base station. This data transfer consumes most of the sensors energy. Therefore, several routing approaches have been proposed to conserve their energy [1]. Most routing protocols designed for small and medium size sensor network provide good performance. However, when the number of nodes increases, traffic control dominates the real communication. This leads to an increase in latency and the explosion of routing tables. To overcome these limitations, routing protocols with hierarchical topology were introduced. These protocols allow the reduction in the number of transmitted messages throughout the network, and thus reduce the energy consumption [2], [3], [4]. In addition, transmission of data to the base station via a single-hop becomes impossible when the size of the network increases. To resolve this issue, the multi-hop communication mode is used to exchange the sensed data among network nodes [4]. Motivated by these reasons, we propose LMEEC, a new multi-hop energy efficient clustering protocol which provides a new way to reduce the sensors energy consumption. In the next section we explain how our protocol addresses these challenges.

## II. LMEEC PROTOCOL

In order to provide flexibility in data routing through the network, LMEEC introduces a layered topology for the network nodes according to the number of hops each of them take to reach the base station. Thus, to achieve high energy efficiency and increase the network scalability, sensor nodes are organized into clusters. For this purpose, we define a new mechanism for grouping nodes into clusters. This mechanism ensures a distribution of the workload of sensor nodes by structuring them into clusters of unequal size. Then, cluster-heads communicate the collected data of the network to the base station. The cluster-heads are selected periodically according to the weight. This weight is calculated so that the number of cluster-heads increases while approaching the base station. Hence, clusters farther away from the base station will have smaller sizes. The execution of LMEEC is established periodically over three phases. The first is the network configuration, while the second ensures the election of cluster-heads and cluster formation. Data communication is the third phase of our protocol.

### A. The Network Configuration

This phase aims to organize the network nodes in layers and discover their neighbors. The neighbors' discovery is launched by the base station using the diffusion of the Hello messages. In doing so, the neighbors at one single-hop of the base station which constitute the first layer of the network hierarchy, are discovered. Thereafter, these nodes will act as the local base station for other neighboring nodes. They explore the network by re-broadcasting the Hello message to the other nodes within their range, while updating their information in the message. At the end of this phase, each node knows its own distance to the base station in terms of the number of hops.

### B. Cluster-head selection / Cluster formation

A sensor node elects itself as a cluster-head by evaluating a weight function $P_i$, and compared it to a threshold. This threshold is inversely proportional to the node layer number. In order to optimize energy management, this weight function should help to choose the nodes with the highest energy capacity, largest number of neighboring nodes, and which have been less frequently cluster-head. The strong idea behind this function is that the degree of involvement of each parameter

varies with the layer number of each node. The weight P of a node i is given by:

$$P_i = \left(\frac{1}{\alpha - L_{num(i)}} \times \frac{deg_i}{N}\right) + \left(\frac{1}{\beta + L_{num(i)}} \times \frac{E_{res}}{E_{total}}\right) - \left(\gamma \times 1 - \frac{1}{1 + num_{CH}}\right) \quad (1)$$

where $E_{res}$ refers to the node's residual energy, $deg_i$ the number of its neighbors, and $num_{CH}$ represents the number of times selected as cluster-head. N is the total number of sensors in the network. The weighted parameters α, β, and γ take values in the interval [0, 1]. They are calculated based on the layer number of node, represented by the variable $L_{num(i)}$.

Once a node is auto-elected as cluster-head, he must inform its new role in the current cycle to the rest of the nodes. For that, it broadcasts an announcement message containing its identifier as well as a weight (P CH) defined according to its residual energy, its degree and its layer's number :

$$P\ CH = \frac{E_{res}}{deg_i} \times l_{num(i)} \quad (2)$$

At the reception of announcement message, the nodes build lists of cluster-heads. Each non cluster-head node determines its cluster by choosing the cluster-head with the greatest P CH weight. The cluster-head has, consequently, the highest energy power and a minimum of sensors to manage. In case of equality, the cluster-head with a great distance far from the base station is privileged. Thereafter, each node sends a message of cluster membership to the chosen cluster-head. This supervised decision makes it possible to balance the load of cluster-heads by producing clusters with different sizes. The size of a cluster decreases accordingly to layer number of its cluster-head : the farthest the cluster-head, the largest the cluster. The objective is to ensure a smaller number of nodes members and a smaller size to manage for cluster-heads of high layers close to the base station. Finally, each cluster-head have a list of his adjacent cluster-heads, which is used to select a relay cluster-head.

### C. Data Communication

In this phase the sensed data is collected and transmitted in a multi-hop fashion to the base station. To ensure the intra-cluster communication, TDMA protocol is used. The cluster member nodes transmit their sensed data during the time slots allocated by their cluster-head. This data is then aggregated by the cluster-head and sent to the relay node. Then, the data transit among relay nodes until reaching the base station. Furthermore, the multiple access CDMA protocol is used so that multiple nodes can simultaneously send their data.

## III. PERFORMANCE ANALYSIS

To evaluate the performance of our protocol, we use J-Sim [5], an open source simulator. Our reference comparison is the LEACH protocol [2]. We vary the number of sensor nodes from 50 to 400. These homogeneous nodes are deployed randomly within a square area of 100mx100m. In each test, the simulation time is set to 500s. The data packets are generated every 0.2s. In addition, the initial battery energy that each node had is set to 2J.

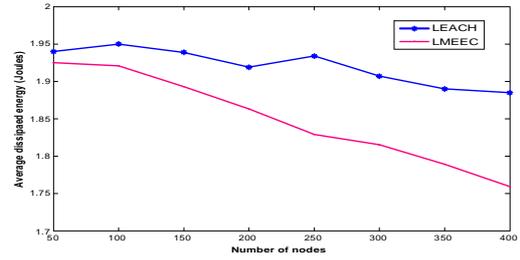
Fig. 1. Average dissipated energy

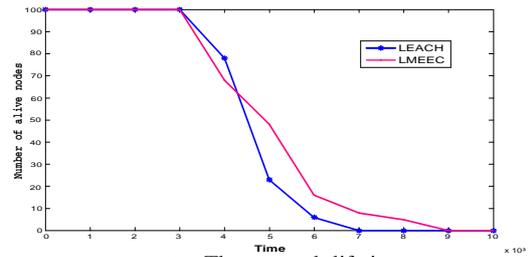
Fig. 2. The network lifetime

Fig. 1 shows the average dissipated energy of LMEEC compared to the results obtained with LEACH. We can observe that, the increasing number of nodes deployed in the network increases the energy consumption in both protocols. Nevertheless, LMEEC consumes less energy; less than 10% of the average consumed energy (for 400 nodes for example) compared to LEACH. This confirms the efficiency of load distribution policy adopted in LMEEC, by taking into account the energy constraint in the selection of cluster-heads nodes and the clusters formation mechanisms. Consequently, LMEEC strongly increases the lifetime of the network compared to LEACH, as shown in Fig. 2. However, it's clear that the lifetime of the network decreases proportionally with the increase of the number of nodes deployed in the network. These results are easily explained by the increasing number of interference between neighboring nodes, when the number of nodes deployed in the network increases.

## IV. CONCLUSION AND FUTURE WORK

We have shown through simulation results that LMEEC outperformed LEACH, in terms of energy efficiency and network lifetime. As future work, we intend to evaluate the generated trafic in the network and study the performance of LMEEC in the presence of mobility.

## V. ACKNOWLEDGMENT

The authors would like to thank Dr. Mubashir Husain Rehmani for his helpful comments and suggestions and Mr. Rafik Zitouni for his interest and support.